\documentclass{emulateapj}
\usepackage{apjfonts}
\newcommand{\scaleup}{\epsscale{1.1}}

\newcommand{\etal}{et al.}

\newcommand{\msun}{M_{\sun}}
\newcommand{\lsun}{L_{\sun}}

\shorttitle{Division Between Quasar \&\ Seyfert Fueling}
\shortauthors{Hopkins \etal}
\slugcomment{Accepted to ApJ, December 14, 2008}
\begin{document}

\title{A Characteristic Division Between the Fueling of Quasars and Seyferts: 
Five Simple Tests}
\author{Philip F. Hopkins\altaffilmark{1}, 
Lars Hernquist\altaffilmark{1}
}
\altaffiltext{1}{Harvard-Smithsonian Center for Astrophysics, 
60 Garden Street, Cambridge, MA 02138}

\begin{abstract}

Given the existence of the $M_{\rm BH}-\sigma$ relation, models of
self-regulated black hole (BH) growth require both a fuel supply and
concomitant growth of the host bulge to deepen the central potential,
or else the system will either starve or immediately self-regulate
without any sustained activity. This leads to a generic prediction
that the brightest quasars must be triggered in major mergers: a large
fraction of the galaxy mass must be added/converted to new bulge mass
and a galactic supply of gas must lose angular momentum in less than a
dynamical time. Low-luminosity active galactic nuclei (AGN), in
contrast, require little bulge growth and small gas supplies, and
could be triggered in more common non-merger events.  This leads to
the expectation of a characteristic transition to merger-induced
fueling around the traditional quasar-Seyfert luminosity divide
(growth of BH masses above/below $\sim10^{7}\,\msun$). We 
compile and survey a number of observations in order to test several 
predictions of such a division, including: (1) A transition to bulge-dominated hosts (which any major
merger remnant, regardless of difficult-to-observe tidal features,
should be). (2) A transition between ``pseudobulges'' and
``classical'' bulges hosting the remnant BHs: pseudobulges are formed
in secular processes and minor mergers, whereas classical bulges are
relics of major mergers. (3) An increase in the amplitude of
small-scale clustering (increased halo occupation of small group
environments) where mergers are more efficient. (4) Different redshift
evolution, with gas-rich merger rates rising to redshifts $z>2$ while
secular processes are relatively constant in time. (5) An increasing
prominence of post-starburst features in more luminous systems. 
Our compilation of observations in each of these areas provides tentative evidence 
for the predicted division around the Seyfert-quasar threshold, 
and we discuss how future observations can 
improve these constraints and, in combination with the tests here, 
break degeneracies between different fueling models. 

\end{abstract}


\section{Introduction}
\label{sec:intro}

The discovery of tight correlations between black hole (BH) mass and
properties of the host galaxy spheroid, including spheroid luminosity
\citep{KormendyRichstone95}, stellar mass \citep{magorrian}, velocity
dispersion \citep{FM00,Gebhardt00}, concentration and profile shape
\citep{graham:concentration,graham:sersic}, have fundamental
implications for the growth of black holes and -- given the
\citet{soltan82} argument which implies that most black hole mass was
assembled in luminous quasar phases \citep[for more recent versions of
this calculation, see
e.g.][]{salucci:bhmf,yutremaine:bhmf,hopkins:bol.qlf,shankar:bol.qlf} --
corresponding quasar activity.

In order to explain not just the normalization but also the small
(factor $\lesssim2$) observed scatter in these correlations
\citep{tremaine:msigma}, models generically require some sort of
self-regulated growth \citep[e.g.][]{silkrees:msigma,
burkertsilk:msigma,wyitheloeb:sam,
dimatteo:msigma,hopkins:lifetimes.methods,
hopkins:bhfp.obs,hopkins:bhfp.theory,
murray:momentum.winds,thompson:rad.pressure}. Because the BH
represents such a small fraction ($\sim10^{-3}$) of the galaxy mass,
it is difficult (if not impossible) to avoid at least the occasional
rare situation in which $\gtrsim100$ times the implied BH mass, from
e.g.\ the $M_{\rm BH}-\sigma$ relation, in cold gas is stripped of
angular momentum and should accrete rapidly (given either e.g.\ Bondi
or thin-disk accretion), giving rise to BHs well off the observed
correlation and orders of magnitude outside the scatter. This
concern has led models to invoke some form of feedback from the BH as
a self-regulation mechanism: regardless of the details, once the BH
reaches some critical mass, if a small fraction ($\lesssim5\%$) of the
radiant energy ($\sim0.5\%$ of the accretion energy) or momentum can
couple to the nearby gas, then simple scalings show that gas will be
unbound and not able to accrete. The equilibrium between the coupled
energy/momentum and the depth of the central potential therefore sets
the maximum black hole mass.

In addition to the nature and small scatter of the observed 
BH-host correlations, there is increasing evidence for a 
process of this nature. \citet{hopkins:bhfp.obs,hopkins:bhfp.theory} pointed out 
that these scenarios generically imply that the 
most basic correlation 
is really between BH mass and central potential depth/spheroid 
binding energy, and demonstrated in the observations 
a $\sim3\,\sigma$ preference for such an association
(i.e.\ the only relationship with no 
residual correlations remaining between BH mass and other host galaxy properties). 
\citet{aller:mbh.esph} found a similar result in an 
independent analysis of the observations modeling the central potential 
in generality, including the contribution of disks and halos. 

Moreover, although it remains unclear which feedback processes
are the most important for this self-regulation, 
numerous such mechanisms are observed to be ubiquitous in 
quasar populations, including intermediate-scale winds, 
accretion disk outflows, jets, 
ionization fronts, Compton heating, 
radiation pressure, and shocks 
\citep[see e.g.][]{weymann:BALs,starkcarlson:m82.nlr,baum:radio.outflow.old,
colbert:seyfert.outflows.1,colbert:seyfert.outflows.2,colbert:seyfert.outflows.3,
laor:warm.absorber,crenshaw:nlr,
elvis:outflow.model,baum:radio.outflows,dekool:large.outflow.1,
cecil:jet.superbubbles, levenson:seyfert.starburst.winds,
ruiz:nlr.kinematics,kaspi:xr.seyfert.kinematics,walter:m82.outflow,
pounds:qso.outflow.1,pounds:qso.outflow.2,biller:agn.hot.gas.outflow,
whittlewilson:jet.ism.interaction,Steenbrugge:outflow.mdot,
rupke:outflows,veilleux:winds,green:qso.outflow.in.lensing,
rice:nlr.kinematics,
gabel:large.outflow,krongold:warm.absorber.outflow.rate,
McKernan:agn.fb.outflow.rates}. 

In addition, although the exact normalization and scatter 
may evolve with redshift \citep[potentially related to evolution in 
host galaxy properties setting the central potential, or 
evolution in feedback physics; see e.g.][]{croton:msigma.evolution,
hopkins:bhfp.theory}, observations increasingly indicate that
BH-host correlations are in place even at high 
redshifts \citep{shields03:msigma.evolution,walter04:z6.msigma.evolution,
merloni:magorrian.evolution,adelbergersteidel:magorrian.evolution,woo06:lowz.msigma.evolution,
peng:magorrian.evolution,shankar:implied.msig.from.gal.ages} and in irregular or disturbed systems 
\citep{borys:xray.ulirgs, alexander:smg.bh.masses,kim:msig.scatter.obs}, 
further demanding some form of self-regulation 
(as opposed to e.g.\ coincidence or fine-tuning from 
non self-regulating BH growth mechanisms). 

If BHs grow in a self-regulated manner, then their evolution 
must be thought of as fundamentally determined by the growth of 
their host spheroid. If the spheroid is not large enough -- 
if the central potential is no deeper -- then a system that has 
already regulated its growth will simply enter into a high-feedback 
mode and immediately self-regulate when new gas is channeled into the 
central regions, {\em regardless of how large a gas supply is fed 
to the BH}. If this were not the case, again the problem of explaining 
the observed tightness of the BH-host correlations would 
present itself. 
{\em ``Growing the monster'' requires not just a fuel supply, but 
also concomitant bulge growth}, in order to raise the limit to which the 
BH can accrete before self-regulating. 

Given the frequency of 
e.g.\ minor mergers and other processes that can remove angular 
momentum from at least some of the gas in the 
central regions of a galaxy \citep{lin:merger.fraction,woods:tidal.triggering,
maller:sph.merger.rates,barton:triggered.sf,
woods:minor.mergers,stewart:mw.minor.accretion,
fakhouri:halo.merger.rates,younger:minor.mergers,hopkins:disk.survival}, simple cosmological calculations (as well as a survey of 
the abundance of massive bulges in the Universe) suggest 
that the {\em bigger} problem which must be overcome in order to 
allow BH growth may be the growth of bulges, {\em not} the removal 
of angular momentum from gas. In any event, the two cannot be 
considered separately.

This consideration leads to a new framework in which Seyfert 
and quasar fueling 
mechanisms must be evaluated. It has long been pointed out that 
major, gas-rich galaxy-galaxy mergers represent a viable 
means to fuel at least some quasars \citep{stockton:3c273.interaction,
heckman84:qso.mergers,
sanders88:quasars,sanders88:warm.ulirgs,
stockton87:qso.fuzz,stocktonridgway:qso.merger.id,hutchingsneff:qso.host.imaging,
sanders96:ulirgs.mergers,bahcall:qso.hosts,
canalizostockton01:postsb.qso.mergers,hutchings:redqso.lowz,
guyon:qso.hosts.ir,dasyra:pg.qso.dynamics, bennert:qso.hosts}, and indeed 
it is generally accepted that although all quasars may not be 
mergers, all gas-rich mergers will induce AGN activity at some level 
\citep[see references above 
and e.g.][]{hopkins:qso.all}. 
These encounters also have the advantage that, when 
two spirals merge, the disks are converted into spheroid and 
the bulge further grows via centrally concentrated star formation 
in a merger-induced starburst \citep{kormendysanders92,mihos:cusps,
mihos:starbursts.94, mihos:starbursts.96, 
hibbard.yun:excess.light, robertson:fp,naab:gas,cox:kinematics,
hopkins:cusps.mergers,
hopkins:cusps.ell,hopkins:cores,hopkins:cusps.fp}. This deepens the 
central potential considerably -- so not only does such a merger 
strip gas of angular momentum at all levels and provide a fuel 
source to the BH, but it also increases
the binding energy of the inflowing material, meaning that the BH 
{\em must} grow larger by a significant factor in order to self-regulate 
and ultimately lie on the observed BH-host correlations. 

There are, however, many alternative mechanisms by which a 
nuclear BH could be fueled, and these will generically be more 
common in gas-rich (especially low-mass) galaxies. These include minor mergers 
(mass ratios $\sim1:10$ or so), angular momentum loss from 
stellar+gas bar systems in disks \citep[for a review see][]{jogee:review}, 
the stochastic accretion of 
cold molecular clouds that happen to be near the BH or 
on scattered trajectories \citep{hopkins:seyferts}, 
and Bondi-Hoyle spherical accretion of hot gas from the 
diffuse, pressure-supported atmosphere in the central 
bulge \citep[see e.g.][and references therein]{allen:jet.bondi.power,best:radio.loudness}. 
However, these mechanisms are much less 
efficient (or even completely inefficient) at growing the 
bulge. Although it is possible to get some growth of the 
bulge and central potential from e.g.\ bars in isolated 
galaxies, most numerical experiments 
and cosmological simulations \citep{combes:pseudobulges,
naab:minor.mergers,bournaud:minor.mergers,
athanassoula:peanuts,debattista:pseudobulges.b,naab:etg.formation,younger:minor.mergers,
hopkins:groups.ell,hopkins:disk.survival} 
as well as observations \citep{oniell:bar.obs,sheth:bar.frac.evol,
jogee:bar.frac.evol,driver:bulge.mfs,
marinova:bar.frac.vs.freq,barazza:bar.colors}
suggest that 
it is becomes difficult to get significant 
growth in these channels above a stellar mass $\sim10^{10}\,\msun$, 
corresponding to a hosted BH mass $\sim10^{7}\,\msun$, and 
the more massive bulge population will be increasingly dominated 
by systems whose growth primarily came in major 
mergers \citep[for more detailed discussion, see][]{kormendy.kennicutt:pseudobulge.review,
maller:sph.merger.rates,hopkins:groups.ell}. This BH mass 
of $\sim 10^{7}\,\msun$ corresponds to a maximum bolometric luminosity 
of $\sim 0.3\times10^{12}\,L_{\sun}$ ($1.3\times10^{45}\,{\rm erg\,s^{-1}}$) 
when radiating at the Eddington limit\footnote{The Eddington 
limit giving $L_{\rm bol} = 3.3\times10^{4}\,L_{\sun}\,(M_{\rm BH}/\msun) = 
1.29\times10^{38}\,{\rm erg\,s^{-1}}\,(M_{\rm BH}/\msun)$.}

Moreover, these distinctions are reinforced by a simple scaling: 
the minimum bolometric luminosity associated with ``true'' quasars 
is $\sim 10^{12}\,\lsun$ \citep{soifer:iras} (this translates to a typical 
optical magnitude $M_{\rm B}<-23$; the traditional Seyfert/quasar 
division introduced by \citet{schmidt:pg.qsos} as an observational, 
not physical division). Fueling a Seyfert ($L_{\rm bol}\le10^{12}\,L_{\sun}$) 
for a \citet{salpeter64} time $t_{S}=4.2\times10^{7}\,{\rm yr}$ (the 
$e$-folding time for exponential growth at the Eddington limit, given a 
typical radiative efficiency expected of BH accretion 
$\epsilon_{r}\equiv L_{\rm bol}/(\dot{M}\,c^{2}) \sim 0.1$), 
similar to the lifetime of a high-accretion rate episode suggested 
by various observational constraints \citep[clustering, 
sizes of radio jets, and the transverse proximity effect; 
for a review see][]{martini04}, requires 
$\lesssim10^{7}\,M_{\sun}$ worth of accreted 
material (unsurprising, given that this luminosity corresponds to a 
BH of around this mass at its Eddington limit). Allowing for 
some reasonable efficiency of fueling (say $\sim10\%$ of the 
mass within the BH radius of influence or vicinity of the BH) this is 
still comparable to the mass in a single or a few giant molecular clouds 
($\sim10^{7}-10^{8}\,\msun$). The limits become less stringent 
as one considers lower luminosities (more typical Seyferts 
with $L_{\rm bol}\sim 10^{11}\,L_{\sun}$ requiring only $\sim10^{6}-10^{7}\,M_{\sun}$ 
worth of gas). There are many processes that could sufficiently 
disturb the gas supply in the central regions of the galaxy so as to produce such 
an event -- for example, even in the Milky Way, a sufficient random perturbation 
(with mass of this magnitude) -- a near pass to the BH of another giant molecular cloud 
complex or massive star cluster -- could torque such a mass (already near the 
BH in molecular clouds) into accretion \citep[see e.g.][and references 
therein]{genzel:gal.center.review}. Given the relative ease of triggering and 
ubiquity of such systems, there is no reason to invoke a rare and violent mechanism 
such as major mergers -- in fact, the natural expectation is that more mundane, 
far more frequent processes should dominate. 

However, 
fueling a bright quasar (say $L_{\rm bol}\sim 10^{14}\,L_{\sun}$ for $M_{B}\approx-28$) 
for a similar duration requires $\gtrsim10^{9}-10^{10}\,M_{\sun}$ in gas -- i.e.\ 
channeling an entire typical galaxy's supply of gas onto the BH itself. 
Given the efficiency of star formation when dense gas is accumulated in the 
central regions of a galaxy, almost any reasonable model demands 
more like $\sim10^{11}\,M_{\sun}$ worth of gas within the central 
$\sim50-100$\,pc (even for objects that are fainter, but still quasars, 
at $L_{\rm bol}\gg 10^{12}\,L_{\sun}$). Given that the timescale needed is 
comparable to or shorter than the dynamical time of the galaxy itself, a 
massive, galaxy-wide perturbation is needed (essentially the entire supply of galactic 
gas must be stripped of angular momentum and free-fall to the center in 
less than a single rotational period of the disk), and the only mechanism 
expected in standard cosmologies to be capable (let alone expected) 
to produce such a perturbation is a major merger (any disk instability or minor 
merger, by the nature of the amplification of that instability or dynamical 
friction, respectively, could neither channel the required mass in an 
absolute sense nor do so in less than a few disk orbital periods). 

This leads to the idea, motivated by 
detailed models for accretion in mergers and isolated systems, 
that there is some characteristic 
host bulge mass/BH mass and corresponding quasar luminosity (which 
happens to correspond to the observationally defined Seyfert-quasar 
division) below which these more ubiquitous mechanisms dominate
AGN fueling (being more common 
and requiring less bulge growth to deepen the central potential 
in this mass regime).  Above this division,  
less violent mechanisms are simply inefficient (they may still happen, but they 
do not sufficiently raise the bulge mass, so BHs  
quickly self-regulate and do not experience any significant 
lifetime of high-Eddington ratio growth) and the population requires 
more extreme mechanisms such as major mergers to build the 
most massive bulges and (corresponding) BHs. 

Here, we outline observational 
constraints that can be
used to test this idea, and provide constraints on 
where non-major merger related fueling mechanisms do or do not 
dominate the BH and AGN populations. These various 
accretion processes have different observational signatures, 
particularly when we recognize that, in a self-regulated 
growth scenario, they must be treated not just as BH growth 
mechanisms but simultaneously grow bulges.
In \S~\ref{sec:morph}-\ref{sec:colors} below, we outline a 
few such tests and illustrate some preliminary 
constraints from previous works, and in \S~\ref{sec:discussion} 
we summarize the results. 

For ease of comparison, we convert all observations 
to bolometric luminosities given the appropriate bolometric corrections 
from \citet{hopkins:bol.qlf} \citep[see also][]{elvis:atlas,richards:seds}. 
We adopt a $\Omega_{\rm M}=0.3$, $\Omega_{\Lambda}=0.7$,
$H_{0}=70\,{\rm km\,s^{-1}\,Mpc^{-1}}$ cosmology and 
a \citet{salpeter:imf} stellar initial mass function (IMF), and normalize all 
observations and models appropriately (note that this generally affects only 
the exact normalization of quantities here, not the qualitative conclusions). 
All magnitudes are in the Vega system. 

\section{Host Galaxy Morphologies}
\label{sec:morph}

A great deal of attention has been paid to the issue of testing merger-induced 
quasar fueling models by looking for signatures of disturbance in 
e.g.\ quasar and AGN hosts \citep{bahcall:qso.hosts,canalizostockton01:postsb.qso.mergers,
floyd:qso.hosts,zakamska:qso.hosts,pierce:morphologies}. However, this is extremely difficult, 
especially at the high redshifts and luminosities of interest. where 
this fueling mechanism is expected to dominate. Moreover, models predict 
that the quasar phase in mergers occurs at the end of the merger, in the 
background of a largely relaxed remnant \citep{dimatteo:msigma,hopkins:lifetimes.letter,
hopkins:lifetimes.methods,hopkins:qso.all}, 
and at these times experiments with e.g.\ automated morphological 
classification systems such as Gini-M20 and CAS \citep{lotz:merger.selection} 
as well as optical identifications \citep{krause:mock.qso.obs} suggest that 
even with perfect image depth, merger induced-quasar hosts would be uniformly 
classified as relaxed, non-merging systems. There is some prospect for 
deeper observations to reveal new insights, as e.g.\ very nearby 
systems (previously classified as relaxed) have, upon deeper
imaging \citep{bennert:qso.hosts} been revealed as clearly interacting, 
and imaging of redder or more obscured quasars has identified 
interaction-dominated populations \citep{hutchings:redqso.lowz,hutchings:redqso.midz,
kawakatu:type1.ulirgs,urrutia:qso.hosts}, with degree of interaction increasing 
in more infrared-luminous populations as well \citep{guyon:qso.hosts.ir}, 
but this remains prohibitively expensive for most systems of interest. 

However, we highlight a more mundane but equally constraining point here. 
Viewed during the final stages of a major merger, it may be impossible to 
(with present observations) say whether 
there was a recent merger/disturbance, but it is possible to identify whether 
the hosts are spheroids or disks. The hosts being spheroids does not 
uniquely imply a merger origin of the hosted quasars, of course, but their 
being disks would clearly put strong constraints on 
merger mechanisms\footnote{Disks can survive mergers 
\citet{barneshernquist96,barnes02:gasdisks.after.mergers,
springel:spiral.in.merger,robertson:disk.formation,
hopkins:disk.survival,hopkins:disk.heating}; but for 
major mergers with large bulge growth (the cases of interest here), 
the remnant disks are generally small and compact and do not 
dominate the stellar mass. Large disks in the remnant require special 
conditions (usually not being coupled to substantial BH fueling); 
so observations of disk-dominated AGN hosts would still present strong 
constraints on what, if any, contribution to fueling comes from major mergers.}.

\begin{figure*}
    \centering
    \scaleup
    %\plotone{host_morphologies.ps}
    \plotone{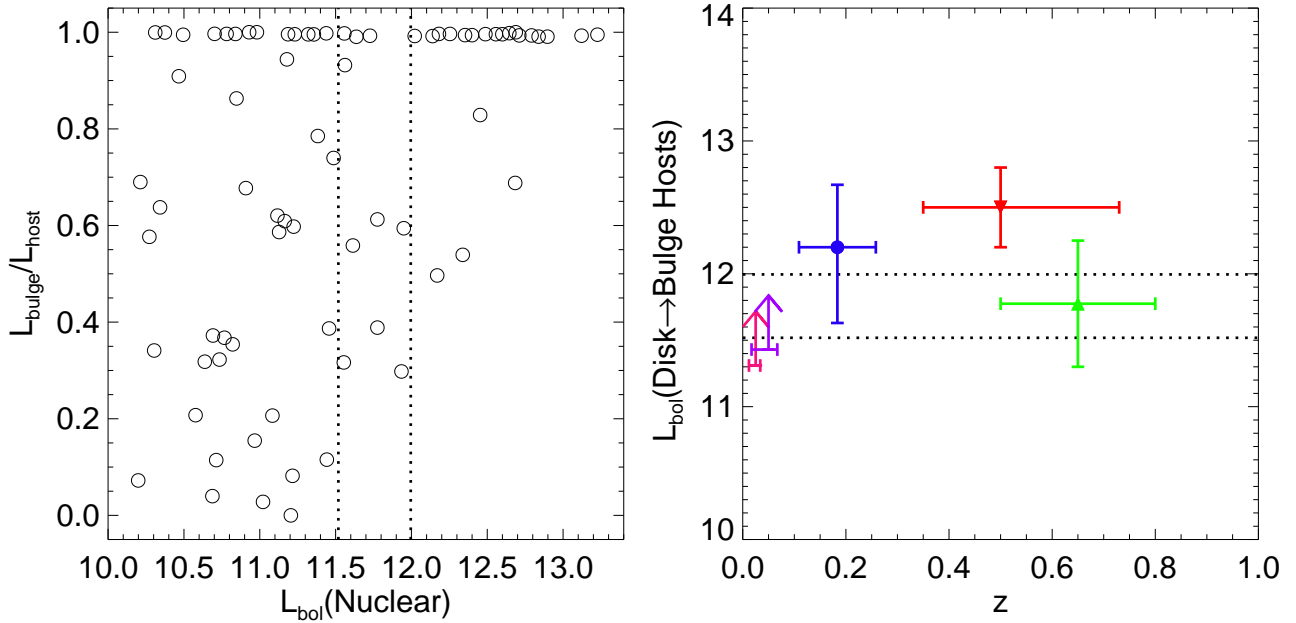}
    \caption{{\em Left:} Host galaxy morphology (bulge-to-total luminosity ratio) 
    versus AGN bolometric luminosity, from the sample of \citet{dunlop:qso.hosts}. 
    Around the characteristic Seyfert-quasar divide ($L_{\rm bol}\sim10^{12}\,L_{\sun}$), 
    hosts transition from being a representative mix of bulges and disks to 
    being exclusively bulge-dominated (Es and S0s): major-merger remnants must be 
    bulge dominated (even if tidal disturbances fade rapidly), whereas non-merger 
    fueling mechanisms allow a random mix of non-spheroid hosts. 
    {\em Right:} Luminosity where the observed AGN host galaxy populations 
    transition from being essentially random or disk-dominated (at low-$L$) to 
    bulge-dominated (at high-$L$), from the observed samples 
    of \citet[][pink arrow]{malkan:qso.host.morph},  
    \citet[][purple arrow]{kauffmann:qso.hosts}, \citet[][blue circle]{dunlop:qso.hosts}, 
    \citet[][red inverted triangle]{zakamska:qso.hosts,zakamska:prep}, 
    and \citet[][green triangle]{rigby:qso.hosts}. The low-$z$ samples do not include bright 
    quasars or see such a transition, so set only a lower limit here. 
    At redshifts $z\lesssim1$ where morphologies are observed over the dynamic range of 
    interest, a characteristic transition in host properties appears roughly at similar 
    luminosity. 
    \label{fig:host.morph}}
\end{figure*}

Indeed, a number of observational studies have pointed out that, among 
the high Eddington-ratio population (important to distinguish from 
possible low-Eddington ratio, non thin-disk accretion in old systems), 
there is a characteristic difference roughly above and below the 
Seyfert-quasar divide. Broad-line Seyferts tend to live in disk-dominated 
host galaxies: clearly, most of these 
systems are not recent major merger remnants. Broad-line 
quasars, on the other hand, have predominately elliptical hosts, candidates 
for recent mergers. 
In fact, \citet{dunlop:qso.hosts,
floyd:qso.hosts} find that the transition regime is 
quite narrow -- around a factor of a few in luminosity, 
where the population 
goes from strongly disk to bulge dominated. 

Figure~\ref{fig:host.morph} presents the results of studies that 
compare the morphological distribution of low and high-luminosity AGN 
selected in the same manner. We show the luminosity below 
which the AGN sample is dominated by either disks or an 
apparently random mix of disks and ellipticals (i.e.\ the sample shows no 
strong preference for early-type hosts) and above which the sample 
is dominated by ellipticals, from the PG quasar 
sample \citep{dunlop:qso.hosts}, the sample of Type II quasar hosts 
in \citet{zakamska:qso.hosts,zakamska:prep}, 
and similar X-ray selected Type II samples \citep{rigby:qso.hosts}. 
Despite significant differences in selection and many of the observed 
properties of the quasars in these samples, the results are similar in each 
study: a sharp transition (over a factor $\sim2$ in $L$) 
between disk/star-forming (or random) host galaxies at 
$L_{\rm bol}\lesssim10^{12}\,L_{\sun}$ and predominantly elliptical or bulge-dominated 
hosts at brighter luminosities.

\section{Relic Bulge Morphologies/Phenotypes}
\label{sec:pseudo}

Numerical simulations and, increasingly, observations have 
established that at least major mergers form ``classical bulges,''\footnote{The 
consequences of minor mergers are more ambiguous. Most minor-merger 
remnants would be classified as ``classical'' bulges down to mergers of 
mass ratios $\sim8:1$ or so, but results are not clear for more 
minor interactions. In minor mergers, the dependence of the 
remnant structure on the initial orbital parameters and other properties 
increases -- for a large range in such parameters, the combination of 
tidal stripping and large orbital times makes it ambiguous whether 
or not the system can even be classified as a 
genuine ``merger.''} whose properties 
resemble those of scaled-down ellipticals (see references 
in \S~\ref{sec:intro}). On the other hand, the 
alternative fueling mechanisms discussed in \S~\ref{sec:intro} 
(bars and sufficiently minor mergers), to the extent that they 
result in any bulge formation at all, yield 
``pseudobulges,'' with distinct properties:
lower Sersic indices and concentrations more analogous to the 
progenitor disk, larger sizes, a higher degree of rotational 
support, more ongoing star formation and bluer colors, and often a 
boxy or ``peanut''-shaped morphology or clear inner disk. 
If there is a characteristic transition regime above and below
which certain 
processes dominate, this should be imprinted in the detailed classes of 
bulges left as relics. 

\begin{figure*}
    \centering
    %\scaleup
    %\plotone{pseudobulge_mfs.ps}
    \plotone{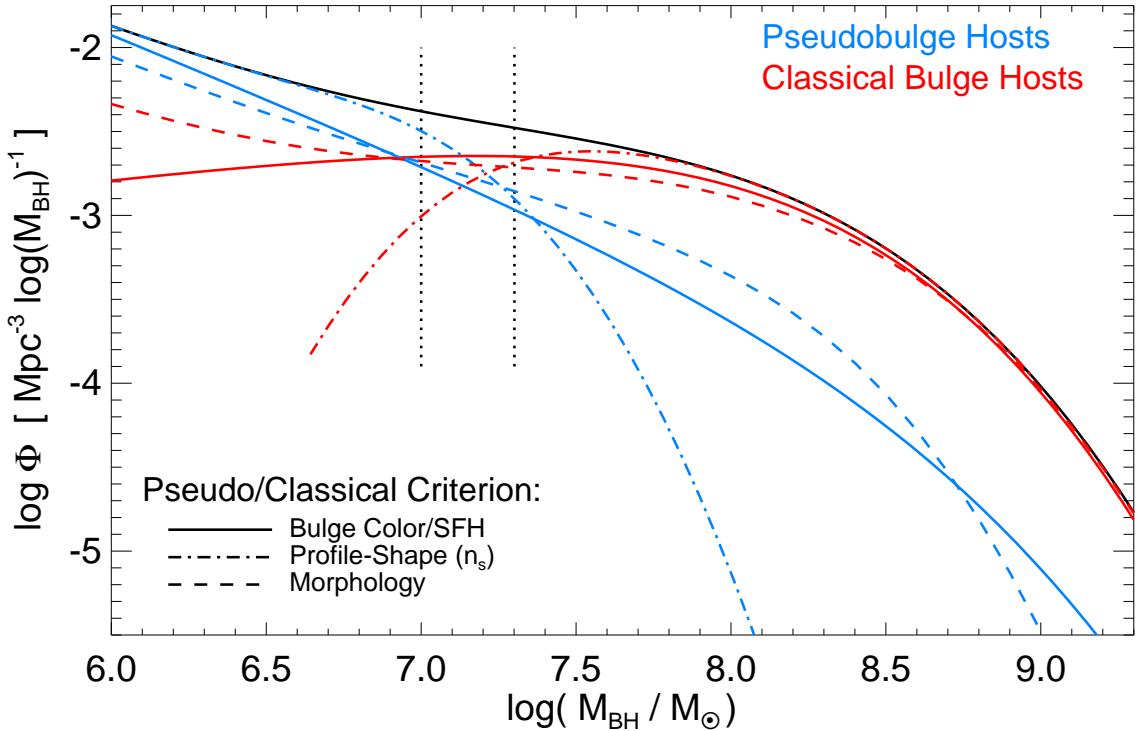}
    \caption{Local BH mass function (black), divided into 
    BHs in ``pseudobulges'' (blue; generally formed in non-merger related secular evolution) 
    and ``classical'' bulges (red; formed in mergers). Different lines adopt different observational 
    criteria for dividing the observed bulge population into pseudo and classical bulges 
    (details in text). Regardless, a characteristic transition occurs at a BH mass 
    $\sim1-3\times10^{7}\,M_{\sun}$, which at the Eddington limit corresponds to the 
    same $\sim10^{12}\,L_{\sun}$ luminosity in Figure~\ref{fig:host.morph}. 
    The BH mass function is consistent with merger-induced fueling above this mass/luminosity 
    limit, and non-merger fueling below the limit. 
    \label{fig:pseudo.mf}}
\end{figure*}

Figure~\ref{fig:pseudo.mf} compares the mass function of BHs in 
local bulges, separated into classical and pseudobulges. 
Given the mass function of bulges/spheroids, we convolve with the 
observed BH-bulge mass relation \citep{marconihunt} and its scatter to 
obtain the corresponding BH mass function. Note that although there 
may be an offset between the BH-host galaxy correlations in 
pseudo and classical bulges, suggested by theory \citep{younger:minor.mergers} 
and observations \citep{hu:msigma.pseudobulges,
greene:pseudobulge.msigma,gadotti:pseudobulge.msigma}, the magnitude of this offset is 
sufficiently small that it makes no difference to our 
comparisons here\footnote{If adopted, this
shifts the mass where the bulge/BH population transitions from 
pseudobulge to classical bulge-dominated to lower mass by 
a factor $\sim1.5-2$, comparable to the existing systematic uncertainties.}. 

Given the spheroid mass function, we classify systems 
as pseudo or classical bulges given one of three different 
(commonly adopted) observational proxies. First, their fitted Sersic indices 
\citep[systems with $n_{s}<2$ being predominantly pseudobulges, 
systems with higher $n_{s}$, similar to observed ellipticals, 
being classical bulges; see][]{fisher:pseudobulge.ns}; we adopt the bivariate 
mass-Sersic index distribution from \citet{driver:bivar.color.concentration.dist,
graham:sersic.bhmf}. Second, their colors 
(bluer-than-average systems being predominantly pseudobulges); 
\citet{drory:quenching.vs.bulge.type} 
and \citet{fisher:pseudobulge.sf.profile} demonstrate a tight correlation of this nature and 
we adopt the corresponding color divisions from \citet{driver:bulge.mfs}. 
Third, their host galaxy types (bulges in systems later 
than Sbc being predominantly pseudobulges), with the type-separated 
mass functions from \citet{kochanek:morph.stellar.mf} and type-dependent $B/T$ values 
from \citet{allerrichstone:bhmf,balcells:bulge.scaling}. 
Regardless of the definition, the estimators 
agree on a similar mass $\sim1-3\,\times 10^{7}\,\msun$ where the transition 
occurs, which at the Eddington limit gives a luminosity 
$\sim10^{12}\,\lsun$ corresponding to the disk-spheroid host division above.

\section{Small-Scale Clustering}
\label{sec:clustering}

The merger rate is enhanced in small-scale overdensities, and this means 
that (with respect to a random, non-merger population with similar masses and 
large-scale clustering properties) recent merger remnants will 
exhibit enhanced clustering on small scales (within of order the size of a 
host halo, $\sim200\,$kpc). This is seen observationally in recent merger 
remnants and post-starburst ellipticals \citep{goto:e+a.merger.connection}, and 
both cosmological simulations and analytic models \citep{scannapieco:sam,thacker:qso.turnover.small.scale.excess,
hopkins:clustering,hopkins:groups.qso} predict that it should leave an observable signature 
in the small-scale clustering of quasars, to the extent that they are 
a merger-induced population. 

\begin{figure*}
    \centering
    \scaleup
    %\plotone{small_scale_clustering.ps}
    \plotone{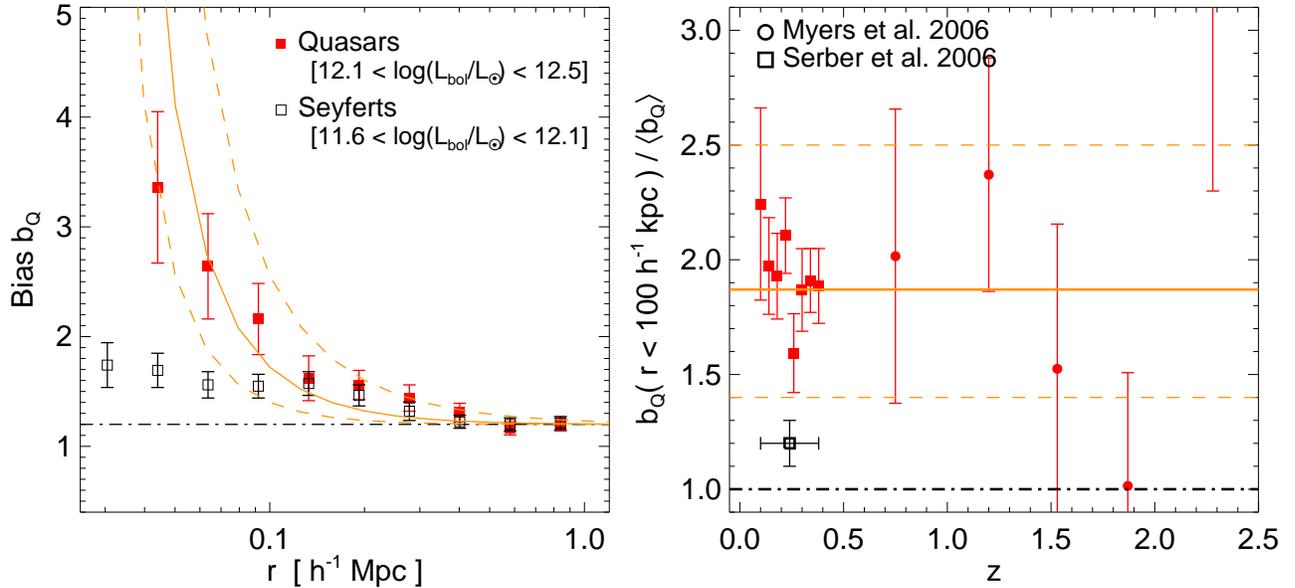}
    \caption{Excess small-scale clustering of quasars and Seyferts. 
    {\em Left:} Observed 
    correlation functions from \citet{serber:qso.small.scale.env} for 
    optical quasars (red filled squares) and Seyferts (black open squares) -- we plot the 
    bias as a function of scale radius determined from the correlation function 
    at each radius relative to that as $r\rightarrow\infty$ (i.e.\ dividing out the best-fit 
    large-scale dark matter correlation function). Dot-dashed line shows a constant bias 
    as a function of scale (i.e.\ no excess of small-scale densities, expected if the 
    galaxies are a random subpopulation with no preference for small groups), 
    solid (dashed) orange line shows the predicted excess on small scales expected 
    for recent merger populations from cosmological models and simulations 
    \citep[reflecting a preference for 
    small group environments; see][]{thacker:qso.turnover.small.scale.excess,
    hopkins:groups.qso}. {\em Right:} Ratio of the mean 
    bias at small radii ($r < 100\,h^{-1}\,{\rm kpc}$) to that at large radii (the asymptotic values in 
    the left panel), at all redshifts where this has been observed. 
    Lines show the predicted excess from 
    the previous panel. We show the same observations from \citet{serber:qso.small.scale.env} 
    (squares), as well as high-redshift observations from \citet{myers:clustering.smallscale} 
    (circles), which include only quasars. Quasars display a similar excess at low and 
    high redshifts (consistent with a constant excess at all $z$), expected in mergers. Seyferts 
    show no such excess at low redshifts, but high-$z$ observations are needed. 
    \label{fig:smallscale.clustering}}
\end{figure*}

Indeed, such a small-scale excess has been seen \citep{hennawi:excess.clustering}, in both 
low-redshift \citep{serber:qso.small.scale.env} and high-redshift 
\citep{myers:clustering.smallscale} quasar populations. 
Interestingly, however, \citet{serber:qso.small.scale.env} see no such excess in the low-redshift 
Seyfert population, with a transition luminosity (between the presence 
and absence of such a feature) corresponding closely to that in 
Figure~\ref{fig:host.morph}. 

We summarize these results in 
Figure~\ref{fig:smallscale.clustering}. At low redshift, we compare the 
relative small-scale bias (amplitude of the observationally estimated 
bias at $r$ relative to that as $r\rightarrow\infty$) for Seyferts and quasars 
in the \citet{serber:qso.small.scale.env} sample \citep[see also][who more robustly describe this 
difference as requiring that quasars have a higher halo occupation in small 
groups]{padmanabhan:qso.smallscale.interp}. The excess seen for quasars (and not for Seyferts) 
is very similar (both qualitatively and quantitatively in magnitude and characteristic 
$r$ where the effect begins) to that seen in known recent merger 
remnants and post-starburst ellipticals \citep{goto:e+a.merger.connection} 
and predicted for recent galaxy merger remnants 
\citep{thacker:qso.turnover.small.scale.excess,hopkins:clustering,hopkins:groups.qso,
hopkins:groups.ell}. 

We can summarize this result by comparing the average bias/clustering 
amplitude within $\sim100$\,pc, relative to that at large $r$ (i.e.\ the 
``excess'' bias on one-halo/group scales), for different samples as a function of 
redshift. Quasar observations extend this (albeit with increasing error bars) 
from $z=0-2.5$, consistent with a relatively constant small-scale clustering excess of a
factor $\sim2$. Unfortunately, comparable observations for Seyferts exist only for 
the low-redshift ($z\lesssim0.5$) sample in \citet{serber:qso.small.scale.env}. This sample is consistent 
with no small-scale excess (i.e.\ requiring that Seyferts show no preference for 
groups or dense small-scale environments, relative to a random galaxy in a halo of 
similar mass), or at least any such excess is smaller than that in quasars at high 
significance, but observations at lower luminosities are required to extend 
this to higher redshifts.

\section{Redshift Evolution}
\label{sec:evolution}

The rates and fractions of gas-rich mergers are expected 
and observed to increase at higher redshift \citep[albeit with debate 
regarding how rapid this increase may be; see e.g.][]{patton:merger.fraction,
conselice:merger.fraction,bundy:merger.fraction,lin:merger.fraction,xu:merger.mf,
depropris:merger.fraction,cassata:merger.fraction,wolf:merger.mf,bundy:mfs,lotz:morphology.evol,
lotz:merger.fraction,bell:merger.fraction,bridge:merger.fractions,kartaltepe:pair.fractions}. 
A truly random subpopulation, on the other hand, will of course have a constant fraction 
\citep[even a gas-fraction dependent population 
driven by e.g.\ random triggering 
through encounters with molecular clouds 
in a gas-rich disk will show similar behavior;][]{hopkins:seyferts}, 
and observations suggest the bar fraction in disks is constant (or even 
decreasing) with redshift \citep{sheth:bar.frac.evol,jogee:bar.frac.evol}. 
Likewise, the fraction of galaxies 
experiencing minor mergers (mass ratios $\gtrsim1:10$) is essentially constant at 
$\sim1$ at all redshifts \citep[not surprising given that the dynamical 
friction time for such a merger is of order the Hubble time;][]{boylankolchin:merger.time.calibration}.

If the average quasar/Seyfert lifetime is (at least in a statistical sense) 
some constant duration or approximately constant fraction of the observable 
merger lifetime -- as suggested by simulations \citep{hopkins:lifetimes.letter,
hopkins:lifetimes.obscuration,hopkins:lifetimes.interp}, mock observations 
\citep{lotz:merger.selection,krause:mock.qso.obs}, and observational constraints on quasar lifetimes and 
duty cycles \citep[see e.g.][and references therein]{martini04} -- then this may be reflected 
in the evolution of the number density or fraction of halos hosting 
AGN of different luminosities (albeit with the caveats in \S~\ref{sec:discussion}).

\begin{figure*}
    \centering
    \scaleup
    %\plotone{downsizing_evolution.ps}
    \plotone{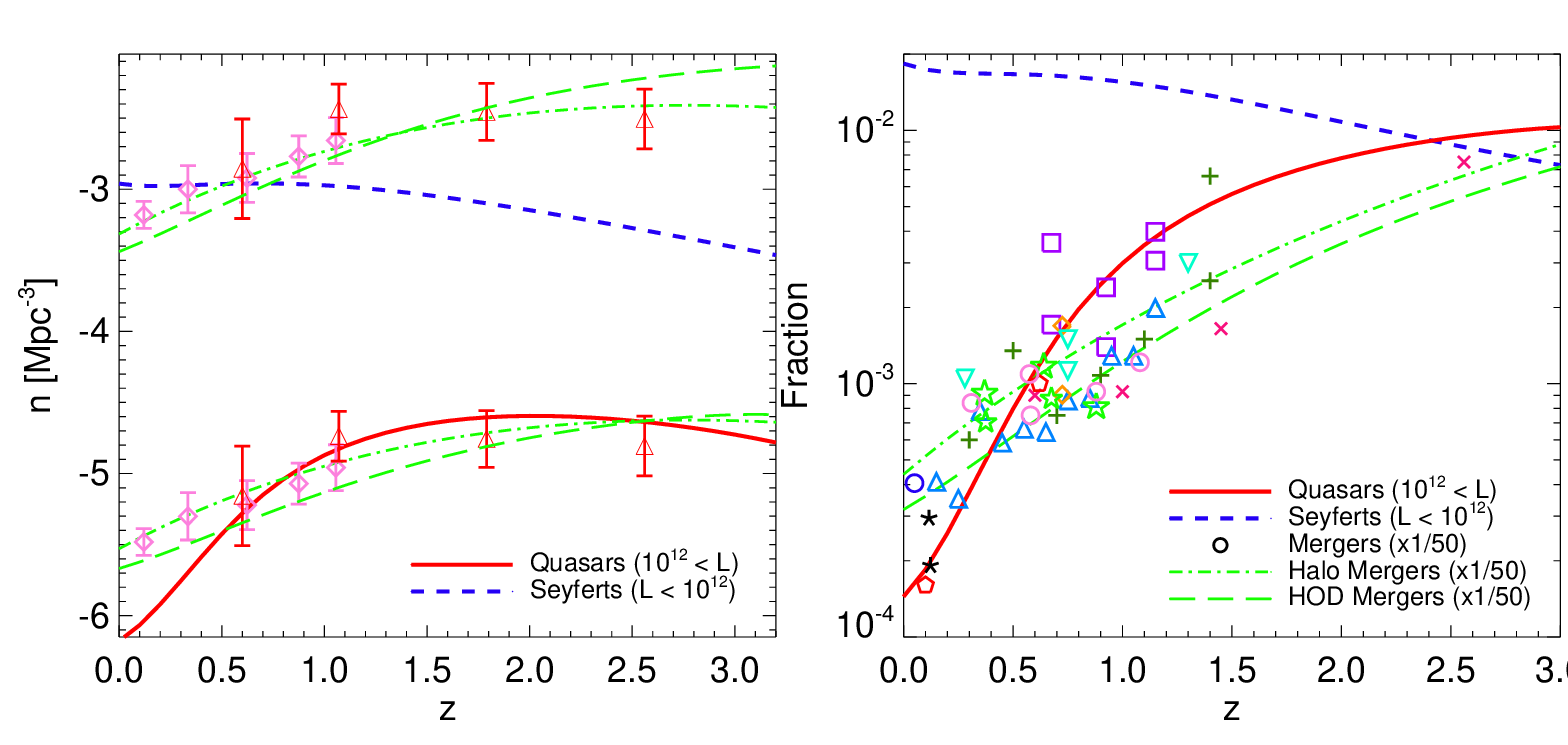}
    \caption{{\em Left:} Observed quasar (red solid) and Seyfert (blue dashed) number densities 
    as a function of redshift, from the compilation of AGN luminosity functions in 
    \citet{hopkins:bol.qlf} \citep[see also][and references therein]{hasinger05:qlf,shankar:bol.qlf}. 
    We compare with the number density of gas-rich (disk-disk) mergers observed 
    \citep[][pink diamonds and 
    magenta triangles with dotted lines, respectively]{lin:mergers.by.type,
    conselice:merger.rate.compilation}, multiplied by 
    $1$ (upper) and $1/50$ (lower), reflecting e.g.\ an approximate ratio of the 
    visible merger timescale ($\gtrsim$\,Gyr) to the quasar/Seyfert lifetime (we 
    are not attempting to constrain the lifetimes here, only the relative redshift evolution). 
    We also compare with predicted merger rates in corresponding 
    dark matter halos 
    \citep[sufficiently massive to host each population; see][green dot-dashed]{fakhouri:halo.merger.rates} 
    and halo occupation models of galaxy-galaxy mergers 
    \citep[also with corresponding mass limits;][green dashed]{hopkins:groups.qso}.
    {\em Right:} Same, in terms of the quasar/Seyfert/merger fraction. AGN fractions are 
    defined in terms of halo occupation models (see text) or the minimum galaxy/BH mass required 
    to host a given AGN. We add merger fractions from 
    %\citet[][filled inverted triangles]{patton:merger.fraction}, 
    \citet[][magenta $\times$'s]{conselice:merger.fraction}, 
    \citet[][green stars]{bundy:merger.fraction,bundy:mfs}, 
    \citet[][pink circles]{lin:merger.fraction}, 
    \citet[][dark blue circles]{xu:merger.mf}, 
    \citet[][black asterisks]{depropris:merger.fraction}, 
    \citet[][cyan inverted triangles]{cassata:merger.fraction}, 
    \citet[][orange diamonds]{wolf:merger.mf}, 
    \citet[][dark green $+$'s]{lotz:morphology.evol,lotz:merger.fraction}, 
    \citet[][red pentagons]{bell:merger.fraction}, 
    \citet[][violet squares]{bridge:merger.fractions}, and 
    \citet[][blue triangles]{kartaltepe:pair.fractions}. 
    Quasar number densities/fractions rise with redshift and peak/flatten at $z>2$, in agreement with 
    merger rates, whereas Seyfert number densities are constant or decline at 
    increasing redshift, more similar to a population with random (constant in redshift) triggering 
    or the observed bar population \citep{sheth:bar.frac.evol,jogee:bar.frac.evol}. 
    \label{fig:downsizing}}
\end{figure*}

Indeed, it is well-established that the number density of bright quasars increases 
rapidly with redshift, peaking at $z\sim2-4$ (above which the population 
must decline, as the absolute number of intermediate to high mass halos/galaxies 
that could host such massive BHs rapidly decreases), whereas the number density of 
low-luminosity Seyferts remains relatively constant, peaking at very low redshifts 
and declining above $z\sim0.5-1$. 

Figure~\ref{fig:downsizing} summarizes these results and compares them to 
expectations of merger rates. First, we compare the absolute number density of 
quasars above the luminosity threshold of interest  
($L_{\rm bol}>10^{12}\,L_{\sun}$) and 
Seyferts ($10^{9}\,L_{\sun}<L_{\rm bol}<10^{12}\,L_{\sun}$). We compare to the 
number density of observed 
mergers\footnote{Technically, we 
are interested in the number of gas-rich mergers, which many of the observations 
considered do not specifically distinguish. However, they are most sensitive to 
these mergers (e.g.\ spiral-spiral mergers), as they are both brighter and tend to 
excite more dramatic tidal features and morphological disturbance than e.g.\ a 
spheroid-spheroid merger of the same mass. Moreover, where observations 
separate the two, the number of ``dry'' or dissipationless spheroid-spheroid 
mergers is much smaller ($\sim 10\%$) than the number of gas-rich mergers 
\citep[e.g.][]{lin:mergers.by.type}.} (with masses $\sim10^{10}-10^{11}\,M_{\sun}$ such that they 
are representative of the galaxy masses hosting these BHs and AGN). 
To lowest order, the number density of merger-induced quasars will scale as 
the merger number density times the ratio of their observable 
lifetimes, $t_{Q}/t_{\rm merger}$. 
Observational selection of mergers is sensitive to the full time of orbital 
decay from some radius (in the case of pair selection) or time since first passage 
(when morphological disturbance is excited) until some number of dynamical 
times after the mergers when features relax; calibration of these selection methods 
against high-resolution hydrodynamic merger simulations suggests 
an observable lifetime $t_{\rm merger}\sim 1-2\,$Gyr. On the other hand, while the 
total time of activity at lower levels may be much longer, 
a typical bright quasar phase lasts only $\sim 20-40\,$Myr 
\citep[see references above or][or considering the time for e.g.\ doubling of 
BH mass in a high Eddington ratio phase]{martini04}. This suggests a value 
of $t_{Q}/t_{\rm merger}\sim1/25-1/50$, which we find provides a good match to the 
relative number densities of quasars and mergers (but in any 
case we are not trying to compare the absolute number of objects, but 
their relative redshift evolution; for our purposes the normalization is arbitrary).

We also compare with the prediction from cosmological models: first, the dark-matter 
only merger fraction (adopting the observable merger timescales from \citet{boylankolchin:merger.time.calibration,hopkins:groups.qso} 
and halo merger rates from \citet{fakhouri:halo.merger.rates}), 
for halos sufficiently massive to host such a
quasar \citep[using the Eddington limit as a lower bound and assuming a fixed 
$M_{\rm BH}-M_{\rm halo}$ relation from][this gives 
$M_{\rm halo}\gtrsim10^{11.5}\,h^{-1}\,M_{\sun}$; 
although the predictions are not sensitive to halo mass]{daangela:clustering}. 
Second, the predicted galaxy merger rate from halo occupation models, fixed to 
reproduce the galaxy mass functions, clustering, and their evolution 
\citep[here from][]{hopkins:groups.ell}. Both are rescaled by the same approximate 
ratio of lifetimes. 

We can reverse this comparison and show the same trends in terms of the 
``active fraction'' -- i.e.\ the fraction of available halos (above a given halo mass or 
hosting a given BH mass) that host a quasar or AGN within some luminosity range. 
This is just the observed merger fraction multiplied by the 
ratio of lifetimes $t_{Q}/t_{\rm merger}$. The quasar fraction in terms of e.g.\ halo 
mass can be defined 
in detail using halo occupation models for quasar clustering 
\citep{croom:clustering,porciani:clustering,fine:mbh-mhalo.clustering,shen:clustering,
daangela:clustering} or simply taking the ratio of the number of quasars to the 
number density of halos sufficiently massive to host any QSO in the relevant luminosity 
bin (both results are identical for our purposes). 
Clearly, regardless of the details of the observable lifetimes, the
quasar number density and fraction rises with redshift in a manner 
roughly consistent with the predicted and observed rates of mergers 
(somewhat faster, in fact, but we have ignored subtleties here such as evolution in quasar 
lifetimes, dependence on host properties, and dependence on the overall gas supply and 
gas content of the merger progenitors).

We repeat these experiments for the Seyfert population (in this case rescaling by a 
ratio of lifetimes $t_{\rm Seyfert}/t_{\rm merger}\sim1$)\footnote{Note that 
we are here considering all Seyferts down to very low AGN luminosities 
$L_{\rm bol}\sim 10^{9}\,\msun$ ($M_{B}=-15$ or $L_{2-10\,{\rm keV}}\sim10^{41.5}\,{\rm 
erg\,s^{-1}}$; including obscured sources at these luminosities) -- at these 
luminosities expected lifetimes are large, $\sim$Gyr (duty cycles are observationally 
seen to be $\sim10\%$), so this ratio $t_{\rm Seyfert}(>10^{9}\,\lsun)/t_{\rm merger}\sim1$ 
is expected.}. 
These trends, regardless of normalization, 
cannot match the observed redshift dependence of Seyfert number density or fractions. 
On the other hand, the weakly evolving (or even decreasing with redshift) observed 
bar fractions can provide a very plausible match, given an average 
ratio of lifetimes $\sim0.3-2$, as can an essentially constant $\sim1$ fraction of 
systems undergoing a minor merger, or the fraction of disks expected to be undergoing 
stochastic activity triggered by random crossings of molecular clouds with the 
BH radius of influence \citep[see][]{hopkins:seyferts}.

\section{Host Galaxy Colors and Star Formation Histories}
\label{sec:colors}

Finally, recent merger remnants should exhibit 
different host galaxy colors and star formation properties 
relative to a random sub-population of 
``normal'' galaxies (or bar/minor merger remnants). 
These characteristics are difficult to measure in practice, 
especially since whatever merger-induced starburst occurs will generally 
leave a dissipative, compact stellar population \citep[effective radii of 
the young populations often $\lesssim1-2\,$kpc in 
mergers and remnants;][]{scoville86,sargent87,kormendysanders92,
hibbard.yun:excess.light, Genzel01,
tacconi:ulirgs.sb.profiles,rj:profiles,rothberg.joseph:kinematics}, 
which could be masked by the quasar PSF and 
will rarely dominate the bolometric light output of luminous QSOs.  

That being said, there has been tremendous effort to constrain the 
stellar populations of AGN observed at many different luminosities and 
in different wavelengths \citep[e.g.][]{brotherton99:postsb.qso,
canalizostockton01:postsb.qso.mergers,
farrah:qso.vs.sf.sed.fitting,farrah:qso.seds.in.warm.ulirgs,
kauffmann:qso.hosts,yip:qso.eigenspectra,
jahnke:qso.host.sf,jahnke:qso.host.uv,
sanchez:qso.host.colors,
vandenberk:qso.spectral.decomposition,barthel:qso.host.sf,
zakamska:qso.hosts}. From these 
efforts, it has been established 
that AGN and star formation activity are broadly contemporaneous. 
This alone does not distinguish between different 
fueling models, so long as a reasonable gas supply is available. 
However, the manner in which the two accompany one another does 
distinguish between models.

A star-forming (disk) galaxy that experiences 
a minor (Seyfert-level) triggering event, be it a minor merger, bar or disk instability, 
or stochastic interaction of BH and giant molecular cloud, will continue (after the AGN 
episode) as just as much a star forming system as before. 
This is true even in the most extreme feedback models, as it 
is nearly impossible for an AGN episode to influence star formation in a 
dense, self-shielding galactic disk with a very large mass 
\citep[see e.g.\ the simulations and analytic calculations in][]{murray:momentum.winds,
thompson:rad.pressure,hopkins:groups.ell}. 
Moreover, it is physically not possible to channel the entire 
galactic gas supply into the center of the galaxy (to be consumed in a 
starburst) in such events (by definition not so in random events or 
minor mergers, and by various conservation laws in disk instabilities). 

On the other hand, a major merger can efficiently exhaust the galactic 
gas supply by channeling gas into nuclear 
starbursts \citep{mihos:starbursts.94,mihos:starbursts.96,
cox:feedback,cox:massratio.starbursts}, leading to associated star formation and 
AGN fueling with a long tail of AGN activity after the final 
coalescence and rapid decline in the star formation rate owing to 
exhaustion of the nuclear gas \citep{springel:models,springel:red.galaxies}.
In short, much (although not 
all) of the merger-induced AGN activity will come after the system has already 
begun to redden as a post-starburst system.  

This is supported by a large number of observations of 
stellar populations in AGN. High Eddington ratio systems 
are observed to be preferentially in bluer-than-average spheroids, 
but systems that are still redder than typical disks of the same 
mass -- i.e.\ spheroid-dominated systems in the so-called ``green 
valley''\footnote{The region in the color-magnitude or color-mass diagram 
\citep[see e.g.][]{strateva:color.bimodality,baldry:bimodality} between 
the locus of star-forming galaxies (the ``blue cloud'') and passively evolving 
spheroids (the ``red sequence'').}
\citep[see e.g.][]{brotherton99:postsb.qso,
kauffmann:qso.hosts,sanchez:qso.host.colors,kewley:agn.host.sf,
nandra:qso.host.colors,silverman:qso.hosts}. 

In particular, \citet{vandenberk:qso.spectral.decomposition} 
argue that brighter quasars (as opposed to 
Seyferts) preferentially show post-starburst activity (rather than a strong 
preference for large amounts of on-going star formation), as expected 
in efficient mergers (and opposite the expectation for secular 
fueling mechanisms), with the strength of these signatures increasing 
in more luminous systems. 
\citet{shi:qso.host.sf.lf} demonstrate that 
the recent star formation activity associated with Seyfert galaxies 
at low redshifts implies that -- although there was some pre-Seyfert 
star formation -- they were sub-LIRG systems 
at the peak of their pre-Seyfert star formation episode (even assuming 
this star formation came in a rapid burst, they would have been 
less luminous than $L\sim 10^{11}\,L_{\sun}$), typical of 
star-forming galaxies. In contrast, the recent star formation 
they identify in quasars is enhanced -- if it came in a sudden 
pre-quasar burst, it would imply these systems were (briefly) 
LIRG/ULIRG systems ($L\gtrsim10^{11}\,\msun$), a luminosity regime 
that is empirically almost universally associated with 
mergers over this redshift range 
\citep{allen85,joseph85,armus87,sanders96:ulirgs.mergers}. 

These luminosity thresholds 
may evolve with redshift, but as a qualitative statement the 
same should be true. At $z=2$, for example, \citet{lutz:qso.host.sf} argue that 
the quasars in their sample require $\sim1000\,M_{\sun}\,{\rm yr^{-1}}$ 
recent star formation episodes \citep[see also][]{wang:highz.qso.ir}, at the hyper-LIRG level 
demanding 
(in almost any model) and ubiquitously associated with major mergers, 
but find that their Seyferts ($L_{\rm bol}\sim3\times10^{11}\,L_{\sun}$) reflect lower-level 
recent star formation \citep[consistent with LIRG-like luminosities, which 
simulations and observations suggest may commonly arise 
from very gas-rich disks at these redshifts without 
major mergers;][]{lefloch:ir.lfs,reddy:z2.lbg.spitzer,
yan:z2.sf.seds,dasyra:z2.sf.imaging,sajina:z2.sf.diagnostics}.

It has been argued that quasars in general exhibit post-starburst signatures 
of a magnitude such that $\gtrsim10\%$ of the stellar mass must have been 
formed in a recent starburst 
\citep[recently see e.g.][for a review see Canalizo \&\ Stockton 2001]{higdon:postsb.qso}. 
Such a large mass fraction is difficult to explain through any 
fueling mechanism other 
than a major merger, and is also in very good agreement with the major-merger 
induced starburst masses implied by observations and 
simulation models of the the surface brightness profiles 
and kinematics of ``classical'' bulges and ellipticals \citep{jk:profiles,hopkins:cusps.ell,
hopkins:cores,hopkins:cusps.fp,hopkins:cusps.evol}. 
The observed kinematics and structural properties of quasar hosts 
are also consistent with this scenario \citep{dasyra:pg.qso.dynamics}, and their location on the 
fundamental plane and associated scaling relations obeyed by classical 
bulges {\em requires} such dissipative star formation in their  
history \citep{schweizer82,LakeDressler86,
hernquist:phasespace,Doyon94,ShierFischer98,James99,
Genzel01,tacconi:ulirgs.sb.profiles,
rj:profiles,rothberg.joseph:kinematics,
dasyra:mass.ratio.conditions,dasyra:pg.qso.dynamics}.

\section{Discussion}
\label{sec:discussion}

In self-regulated scenarios for the growth of black holes, growing the bulge 
(deepening the central potential around the BH) is as much a requirement of 
sustained bright, high Eddington ratio quasar activity as is actually getting 
a fuel supply of cold, low angular momentum gas into the vicinity of the 
BH. Moreover, fueling a Seyfert requires something like the mass of a few 
giant molecular clouds near the BH, whereas fueling a bright quasar requires 
channeling a galaxy's worth of cold gas into the central $\sim10-100$\,pc. 
This generically leads to the expectation that non-merger related 
models for quasar fueling become inefficient around a characteristic 
BH mass of $\sim1-3\times10^{7}\,\msun$ (host bulge mass $\sim10^{10}\,\msun$), 
corresponding to the characteristic bolometric luminosity of 
these BHs near Eddington, $\sim10^{12}\,\lsun$ -- remarkably 
near the traditional Seyfert-quasar divide \citep[$M_{B}\approx-23$;][]{schmidt:pg.qsos}. 
Essentially, this is a restatement of what has become the conventional 
wisdom and is increasingly established by observation in the field of 
elliptical galaxy/bulge formation: massive (predominantly ``classical'') 
bulges are formed in major mergers, whereas low-mass bulges 
(predominantly ``pseudobulges'' in low-mass disks with types 
Sb and later) have large populations formed by non-merger 
mechanisms (i.e.\ bars and instabilities, or bar-like activity triggered 
in sufficiently minor mergers). 

We outline five simple tests to distinguish these modes 
of fueling and 
determine whether (and if so where) such a transition occurs. 
In each case, there is at least tentative evidence for the scenario we 
outline, although more detailed observations as a function of 
AGN luminosity and redshift are needed to develop more 
robust constraints. True quasars tend to have spheroidal hosts, 
and their relic BHs live in classical bulges; they exhibit characteristically 
strong small-scale clustering; their redshift evolution from 
$z\sim0-2$ is similar to that of the gas-rich merger population; 
and their colors indicate their nature as post-starburst systems. 
Seyferts, on the other hand, tend to live in a broader mix of hosts 
or in disk-dominated galaxies, and their relic BHs exist in 
pseudo-bulges; they exhibit weaker small-scale clustering, consistent 
with random galaxies of similar mass; their redshift evolution is 
weak -- their abundance remains constant or even decreases with redshift, 
similar to what is observed in the fraction of barred disks and predicted 
for stochastic (quiescent) fueling mechanisms; their star 
formation histories are suggestive of blue, star-forming galaxies. 

It is important to emphasize that none of these constraints uniquely 
or individually implies e.g.\ a merger origin for bright quasars. 
For example, the morphologies could reflect their simply being 
re-ignited by e.g.\ minor mergers, gas cooling, or stellar mass loss 
in evolved ellipticals; but this is difficult to reconcile with their 
clustering (why they do not trace random, more evolved ellipticals in such a 
case), colors (why they appear post-starburst), or redshift evolution (why 
they would not grow in abundance at late times, as the cooling rates, stellar 
mass loss rates, and number density of such ellipticals are all higher). 
Their classical bulge relics could have been transformed from ``initial'' 
pseudo-bulge hosts at the time of the quasars by dissipationless (dry) 
mergers\footnote{In practice, this is probably not possible without new 
dissipation to effectively re-build the bulge, because conservation of 
phase space densities in dissipationless mergers means that pseudobulges -- 
which characteristically have low central densities corresponding to 
fitted Sersic indices to their mass profiles with $n_{s}\lesssim2$ -- 
cannot be dissipationlessly re-merged and produce something with the 
high central densities characteristic of classical bulges (reflected 
in Sersic indices $n_{s}\gtrsim4$).}, but again this fails to explain their clustering 
and colors, and makes the prediction that there should be very little 
true classical bulge population at intermediate redshifts. 
There are many 
ways to populate galaxies as subhalos that reproduce the 
clustering signature on small scales 
\citep[see the discussion in][]{padmanabhan:qso.smallscale.interp}, 
but those that correspond to a reasonable fueling hypothesis: e.g.\ 
that AGN are random spirals in a certain mass range in groups, or rapidly 
star-forming objects/LIRGs, fail to reproduce the other
 observations considered. Redshift evolution 
and star formation histories are similarly degenerate under various 
fueling models, but again it is the particular combination of 
features that is difficult to reproduce. 

Together, then, these argue for a scenario with a reasonably rapid (and 
not strongly redshift-dependent) transition between major merger 
and non-major merger fueled populations around the 
traditional Seyfert-quasar divide. If true, this has a number of corresponding 
implications for the demographics of BHs and AGN: BHs less massive 
than $\sim1-3\times10^{7}\,M_{\sun}$ (and their corresponding host bulges 
$\lesssim10^{10}\,M_{\sun}$, typically Sb and later-type galaxies) would be 
built up primarily via non-merger mechanisms, whereas more massive black 
holes and bulges -- the ``classical'' systems generally associated with 
early-type galaxies (Sa and earlier) -- would be built up primarily via 
mergers. In an integral sense, then, these merger-built systems dominate 
the total mass density of black holes (with most of the mass density in 
$\sim 10^{8}\,M_{\sun}$ black holes near the break in the black hole 
mass function), constituting $\sim80\%$ of the mass in BHs \citep[see e.g.][]{marconi:bhmf}. 
Correspondingly (by the Soltan 1982 argument), these systems should 
dominate the total luminosity density of quasars and AGN (integrated over 
all redshifts). 
Given the observed evolution of the quasar luminosity function, 
a redshift-independent division at the quasar-Seyfert divide 
implies that the luminosity density is dominated by 
mergers at high redshifts, with a transition to 
dominance of non-merger induced fueling 
below redshifts $z\sim1$. Despite their relative lack of 
importance in the integrated AGN luminosity density, 
because of the sensitivity to lower-redshift AGN, 
this suggests that a large (perhaps dominant) fraction of the X-ray 
background is contributed by these non-merger related systems. 
There are clearly a large number of important implications 
for such a distinction. 

A major caveat to this comparison is that, depending on 
the selection depth, wavelength, and methodology, the Seyfert population 
recovered could have very different properties, biasing it to be 
more similar to or different from the quasar population. 
This is because, unlike at bright quasar luminosities which effectively require 
massive BHs near Eddington, there are two ways to achieve, in practice, 
Seyfert luminosities: a low-mass BH at high Eddington ratio and a 
high-mass BH at low Eddington ratio. 
High mass systems at low Eddington ratio could be the long-lived 
remnants of mergers, slowly decaying in AGN luminosity after 
some initial, bright quasar phase or re-activated by e.g.\ accretion of new 
gas in cooling flows or minor mergers, exhibiting characteristically early-type, 
classical-bulge dominated 
hosts, with post-starburst populations that might be young for 
ellipticals but are older than those seen in spiral galaxies. Small-scale 
clustering signatures in such populations may remain if they were 
merger-triggered, but it depends significantly on their long-term evolution. 
The number density evolution of such systems, being cosmologically 
long-lived and not directly tied to an instantaneous merger rate (even 
if they are merger-triggered) will reflect, as we predict above, a 
relatively constant order unity duty cycle. 
Being at low Eddington ratio, and therefore dim relative to their 
hosts in optical bands (and possibly radiatively inefficient), such 
systems may be more commonly seen in X-ray or narrow-line 
surveys. 

On the other hand, surveys of broad-line or optical/IR dominant 
Seyferts may be more likely to pick out systems at high Eddington ratio, 
i.e.\ low-mass, gas-rich systems that might be triggered by 
minor mergers, bar or disk instabilities, or stochastic encounters 
with molecular clouds. These systems will be disky, gas-rich, and 
pseudobulge-dominated, with no small-scale clustering preference, 
a constant order unity duty cycle, and more ongoing star formation. 
These distinctions are discussed in detail in \citet{hopkins:seyfert.bimodality}; 
here, we have attempted to develop a robust set of predictions that, 
together, can distinguish AGN populations connected to mergers 
from either of these ``quiescent'' modes of fueling Seyferts (or from 
the most likely Seyfert population, some mix of the two). But in several 
of the tests proposed here, these different fueling modes at low luminosities 
will manifest differently, and thus they should be considered means to 
constrain and discriminate among different low-luminosity populations, 
as well.

\acknowledgments We thank Mike Brotherton, Josh Younger, and Pat Hall 
for helpful conversations and discussion in the development of this paper. 
This work 
was supported in part by NSF grants ACI 96-19019, AST 00-71019, AST 02-06299, 
and AST 03-07690, and NASA ATP grants NAG5-12140, NAG5-13292, and NAG5-13381.

\bibliography{/Users/phopkins/Documents/lars_galaxies/papers/ms}

\end{document}